\documentclass[superscriptaddress,twocolumn,showpacs,preprintnumbers,amsmath,amssymb]{revtex4}

\usepackage{graphicx}
\usepackage{dcolumn}
\usepackage{bm}
\begin{document}


\title{Fluctuation and dissipation dynamics in fusion reactions\\ from stochastic mean-field approach}

\author{Sakir Ayik}
\affiliation{Physics Department, Tennessee Technological University, Cookeville, TN 38505, USA}

\author{Kouhei Washiyama}
\affiliation{GANIL, Bd Henri Becquerel, BP 55027, 14076 Caen Cedex 5, France}

\author{Denis Lacroix}
\affiliation{GANIL, Bd Henri Becquerel, BP 55027, 14076 Caen Cedex 5, France}

\date{\today}

\begin{abstract}
By projecting the stochastic mean-field dynamics on a suitable
collective path during the entrance channel of heavy-ion collisions,
expressions for transport coefficients associated with relative distance are extracted. 
These transport coefficients, which have similar forms to those familiar from
nucleon exchange model, are evaluated by carrying out TDHF simulations. 
The calculations provide an accurate description of the
magnitude and form factor of transport coefficients associated with
one-body dissipation and fluctuation mechanism.
\end{abstract}


\maketitle

\section{INTRODUCTION}

The self-consistent mean-field theory, also known as time-dependent
Hartree-Fock (TDHF), by employing Skyrme-type effective interactions, has
been extensively applied to describe nuclear collision dynamics at low
bombarding energies below 10~MeV per nucleon~\cite{Ring,Goeke,Davis,Sim08}. 
In the mean-field theory, short range two-body correlations are neglected and 
nucleons move in 
the self-consistent potential produced by all other nucleons. This is a good
approximation at low energies since Pauli blocking is very effective for
scattering into unoccupied states. Consequently, in the mean-field theory,
collective energy is converted into intrinsic degrees of freedom via
interaction of nucleons with the self-consistent mean-field, so-called
one-body dissipation~\cite{Koonin,Negele}. One-body dissipation mechanism 
plays dominant role in low energy nuclear dynamics including deep-inelastic heavy-ion
collisions and heavy-ion fusion reactions. One important limitation of the
mean-field theory is related with dynamical fluctuations of collective
motion. In the mean-field description, while single-particle motion is
treated in quantal framework, collective motion is treated almost in
classical approximation. Therefore, TDHF provides a good description for 
average evolution, however it severely underestimates
fluctuations of collective variables.

On the other hand, it is well known that no dissipation takes place without
fluctuations~\cite{Gardener,Weiss}. Much effort has been done to improve 
one-body transport
description beyond the mean-field. Most of these transport descriptions take
into account dissipation and fluctuation mechanisms due to two-body collisions,
which play an important role in nuclear dynamics 
at intermediate energies~\cite{Ayik1,Randrup1,Abe,Lac04}. 
Here, we deal with nuclear dynamics at low energies at which
one-body dissipation and associated mean-field fluctuations play a
dominant role in dynamical evolution. One of the fundamental questions is
how to improve the mean-field theory by incorporating one-body
fluctuation mechanism at a microscopic level? In a recent work, based on an
appealing idea of Dasso~\cite{Dasso1,Dasso2}, this question has been addressed. A
stochastic mean-field (SMF) approach has been proposed for describing
fluctuation dynamics~\cite{Ayik2,Ayik3}. For small amplitude fluctuations, this model
gives a result for dispersion of a one-body observable that is identical to
the result obtained through a variational approach~\cite{Bal84}. It is also shown that,
when the SMF is projected on a collective variable, it gives rise to a
generalized Langevin equation~\cite{Mori}, which incorporates one-body dissipation and
one-body fluctuation mechanisms in accordance with quantal
dissipation-fluctuation relation. These illustrations give a strong
support that the SMF approach provides a consistent microscopic description
for dynamics of density fluctuations in low energy nuclear reactions. In
this paper, we present another demonstration of the SMF approach.

In a recent work, by a suitable definition of collective variables of
relative motion, the nucleus-nucleus potential energy and one-body friction
coefficient as a function of relative distance have been extracted from
simulations of microscopic TDHF~\cite{Washiyama1}, see also~\cite{Umar}. 
Such a reduction is not
constrained by adiabatic or diabatic approximation, therefore it should
provide an accurate description of conservative nucleus-nucleus potential
energy and the magnitude of the one-body dissipation mechanism~\cite{Washiyama2}. 
It is of great interest to deduce magnitude of diffusion coefficients associated
with collective variables. However, this information is not contained in the
standard mean-field approximation. The SMF approach provides a proper
framework for extracting dissipation and fluctuation properties of
collective variables. In this work, we carry out a similar macroscopic
reduction of the SMF approach on a collective path. In this manner, we
deduce not only nucleus-nucleus potential and one-body friction coefficient,
but also one-body diffusion coefficients associated with collective
variables.

In Sec.~II, we give a brief description of the SMF approach. In Sec.~III,
we present a suitable definition of collective variables in heavy-ion collisions,  
and the correlation function of Wigner distribution. In Sec.~IV, 
we derive transport coefficients associated with 
relative motion from the SMF approach. In Sec.~V, conclusions are given.

\section{STOCHASTIC MEAN-FIELD APPROACH}

In the standard TDHF, temporal evolution of the system is described by a
single Slater determinant constructed with time-dependent single-particle
wave functions $\Phi _{j \sigma\tau} (\boldsymbol {r},t)$. Evolution of single-particle wave
functions are determined by the TDHF equations with proper initial conditions,
\begin{eqnarray}
\label{eq1}
i\hbar \frac{\partial }{\partial t}\Phi _{j \sigma\tau} (\boldsymbol {r},t) 
= h(\rho )\Phi _{j \sigma\tau}(\boldsymbol {r},t),
\end{eqnarray}
where $h(\rho )$ denotes the self-consistent mean-field Hamiltonian
with the one-body density $\rho$. 
For clarity of notation spin-isospin quantum numbers 
$\tau=($proton, neutron$)$ and $\sigma=($spin-up, spin-down$)$ are explicitly
indicated in these expressions. In many
situations, it is more appropriate to express the mean-field approximation
in terms of the single-particle density matrix,
\begin{eqnarray}
\label{eq2}
\rho (\boldsymbol {r},\boldsymbol {r}^{\prime},t) = \sum\limits_{j \sigma\tau} 
\Phi _{i \sigma\tau}^\ast (\boldsymbol {r},t)n_j^{\sigma\tau}
\Phi _{j \sigma\tau} (\boldsymbol {r}^{\prime},t),
\end{eqnarray}
where $n_j^{\sigma\tau}$ denotes occupation factors of single-particle states.
In the standard TDHF, occupation factors are one and zero for the occupied 
and unoccupied states, respectively. If the initial state is at a finite temperature, the average
occupation factors are determined by the Fermi-Dirac distribution.

TDHF provides a deterministic evolution of the single-particle
density matrix, starting from a well-defined initial state and leading to a
well-defined final state. In order to incorporate fluctuation mechanism into
dynamics, we give up standard description in terms of a single Slater
determinant, and consider a superposition of determinantal wave functions.
As a result of correlations, initial density cannot have a deterministic
shape, but it must exhibit quantal zero-point fluctuations, and if the
initial state is at a finite temperature, it also involves thermal
fluctuations. In the SMF approach the initial density fluctuations are
incorporated into the description in a stochastic manner~\cite{Ayik2}. 
The initial
density fluctuations are simulated by representing the initial state in
terms of a suitable ensemble. In this manner, an ensemble of density
matrices is generated,
\begin{eqnarray}
\label{eq3}
\rho ^\lambda (\boldsymbol {r},{\boldsymbol {r}}^{\prime},t) = \sum\limits_{ij \sigma\tau} 
\Phi_{i \sigma\tau}^\ast(\boldsymbol {r},t;\lambda )\rho_{ij}^\lambda(\sigma\tau) 
\Phi_{j \sigma\tau} (\boldsymbol {r}^{\prime},t;\lambda ).
\end{eqnarray}
Here $\Phi _{j \sigma\tau} (\boldsymbol {r},t;\lambda )$ is a complete set of 
single particle basis,
$\lambda $ denotes a member in the ensemble, and matrix elements 
$\rho_{ij}^\lambda(\sigma\tau) $ are time-independent random Gaussian numbers. Gaussian
distribution of each matrix element is specified by a mean value
$\overline {\rho _{ij}^\lambda}(\sigma\tau)=\delta _{ij}n_j^{\sigma\tau}$, 
and a variance,
\begin{eqnarray}
\label{eq4}
&&\overline {\rho _{ij}^\lambda(\sigma\tau)
\rho _{{j}'{i}'}^\lambda( {\sigma}'{\tau}')} \nonumber\\
&=& \frac{1}{2}
\delta_{j{j}'}\delta _{i{i}'}\delta _{\tau {\tau}'} \delta _{\sigma {\sigma}'}
\left[n_i^{\sigma\tau}(1 - n_j^{\sigma\tau}) + n_j^{\sigma\tau}(1 - n_i^{\sigma\tau})\right],
\nonumber\\
\end{eqnarray}
where $n_j^{\sigma\tau}$ denotes the average occupation factor for a given values of
spin-isospin quantum number $\sigma$ and $\tau$.  
$\delta _{\tau {\tau}'} $ and $\delta _{\sigma {\sigma}'} $
indicate that density matrix elements are assumed to be uncorrelated 
in spin-isospin space. A member of the
ensemble is generated by evolving the single-particle wave functions
according to the self-consistent mean-field of that member,
\begin{eqnarray}
\label{eq5}
i\hbar \frac{\partial }{\partial t}\Phi _{j \sigma\tau} (\boldsymbol {r},t;\lambda ) 
= h(\rho^\lambda )\Phi _{j \sigma\tau} (\boldsymbol {r},t;\lambda ),
\end{eqnarray}
where $h(\rho ^\lambda )$ is the self-consistent mean-field Hamiltonian in that event.

\section{STOCHASTIC WIGNER DISTRIBUTION}

In order to carry out projection of the SMF on a collective space, to determine
transport coefficients of collective variables and to establish connection with
the collective transport models, it is very convenient to introduce the
stochastic Wigner distribution. The Wigner distribution for each event $\lambda $ is
defined as a partial Fourier transform of density matrix as
\begin{eqnarray}
\label{eq6}
&&f^\lambda (\boldsymbol {r},\boldsymbol {p},t)= \int d^3s\exp \left( - \frac{i}{\hbar } 
\boldsymbol{p} \cdot \boldsymbol {s}\right)\nonumber\\
&\times&\sum\limits_{ij \sigma\tau} 
\Phi _{j \sigma\tau}^\ast \left(\boldsymbol {r} + \frac{\boldsymbol{s}}{2} ,t;\lambda \right)
\rho _{ji}^\lambda(\sigma\tau) 
\Phi _{i \sigma\tau}\left(\boldsymbol {r} - \frac{\boldsymbol{s}}{2},t;\lambda \right).
\end{eqnarray}

\begin{figure}[tbhp]
\begin{center}\leavevmode
\includegraphics[width=\linewidth, clip]{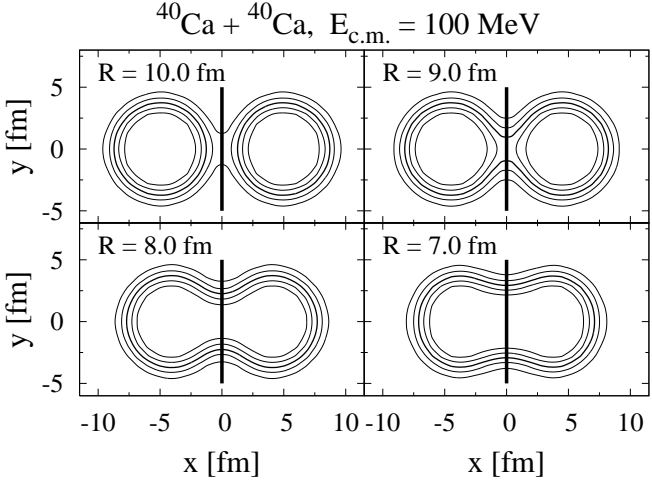}
\caption{
Density profiles $\rho(x,y,0)$ obtained with TDHF for 
the $^{40}$Ca${}+^{40}$Ca reaction at $E_{\rm c.m.}=100$~MeV
at different $R$. The iso-densities are plotted every 0.025 fm$^{-3}$.
In each case, the vertical line indicates the separation plane. 
}
\label{fig:density}
\end{center}
\end{figure}

In this work, we focus on head-on collisions of two heavy-ions and take the
collision direction as the $x$-axis. Following Ref.~\cite{Washiyama1}, 
we define center-of-mass coordinate $R_\pm^\lambda $, total momentum $P_\pm^\lambda $ and
mass number $A_\pm^\lambda $ of projectile-like ($+$) and target-like ($-$)
fragments by introducing the separation plane. The separation plane can be
conveniently defined as the plane at position where iso-contours of
projectile-like and target-like densities cross each other. 
We indicate position of the separation plane, i.e., position of the window at $x = x_0 $.
Illustration of density profiles and separation plane locations are displayed at different times
of the symmetric reaction $^{40}$Ca${}+^{40}$Ca in Fig.~\ref{fig:density}. For calculations
in this figure and the rest of the paper, we use three-dimensional TDHF code developed by P. Bonche and
co-workers with the SLy4d Skyrme effective force~\cite{kim97} and for technical details please see
Ref.~\cite{Washiyama1}.  

It is convenient to express macroscopic variables in each event in terms of
the reduced Wigner distribution $f^\lambda (x,p_x ,t)$ according to
\begin{eqnarray}
\label{eq7}
R_ \pm ^\lambda =\frac{1}{A_{\pm}^{\lambda}} 
\int {\frac{dxdp_x }{2\pi \hbar }}\theta (x_0 \pm x) x f^\lambda (x,p_x ,t),
\end{eqnarray}
\begin{eqnarray}
\label{eq8}
P_ \pm ^\lambda = \int {\frac{dxdp_x }{2\pi \hbar }}\theta (x_0 \pm x)p_x f^\lambda (x,p_x ,t),
\end{eqnarray}
and
\begin{eqnarray}
\label{eq9}
A_\pm^\lambda=\int\frac{dxdp_x}{2\pi \hbar}\theta (x_0 \pm x)f^\lambda(x,p_x ,t).
\end{eqnarray}
We note that these definitions do not involve semi-classical approximations
and are fully equivalent to those given in~\cite{Washiyama1}. 
The ratio $P_ \mp ^\lambda /
\dot {R}_ \mp ^\lambda = m_ \mp ^\lambda (R)$ determines inertia of both
sides of the window and the relative momentum is defined as
\begin{eqnarray}
\label{eq10}
P^\lambda = \frac{m_ - ^\lambda  P_ + ^\lambda - m_ + ^\lambda  P_ - ^\lambda
}{m_ - ^\lambda + m_ + ^\lambda } = \mu ^\lambda (R)\dot {R}^\lambda,
\end{eqnarray}
where $\mu ^\lambda (R) = m_ + ^\lambda m_ - ^\lambda /
(m_ + ^\lambda + m_ - ^\lambda )$ and $\dot {R}^\lambda = \dot {R}_ +
^\lambda - \dot {R}_ - ^\lambda $ are the reduced mass and the
relative velocity of projectile and target sides, respectively. The
reduced Wigner distribution $f^{\lambda} (x,p_x ,t)$ is obtained by
integrating over the phase-space variables $y,z,p_y ,p_z $ according to
\begin{eqnarray}
\label{eq11} f^{\lambda} (x,p_x ,t) = \int\!\!\!\int 
{dydz\frac{dp_ydp_z }{(2\pi \hbar )^2}} f^{\lambda} 
(\boldsymbol{r},\boldsymbol{p} ,t).
\end{eqnarray}

In order to extract diffusion coefficients associated with collective
variables, we need different-time correlation function of the reduced
Wigner distribution on the window. Assuming that the amplitude of density
fluctuations is small, this correlation function on the window is calculated in the
semi-classical approximation in Appendix A to give
\begin{eqnarray}
\label{eq12}
&&\overline {\delta f^\lambda (x,p_x ,t)\delta f^\lambda (x,{p}'_x ,{t}')}
\vert _{x =x_{0}} \nonumber\\
&=& (2\pi \hbar )\frac{m}{\vert p_x \vert }\delta (p_x - {p}'_x
)\delta (t - {t}')\Lambda^{+}(x_{0},p_x ,t),
\end{eqnarray}
where the quantity $\Lambda^{\pm}(x_0,p_x ,t)$ is defined as
\begin{eqnarray}
\label{eq13}
\Lambda^{\pm}(x_0,p_x ,t)&=&\sum\limits_{\sigma\tau} \{f_P^{\sigma\tau}(x_0,p_x ,t)
\left[1 - \bar{f}_T^{\sigma\tau}(x_0,p_x ,t)\right] \nonumber\\
&&{}\pm f_T^{\sigma\tau}(x_0,p_x ,t)\left[1 - \bar{f}_P^{\sigma\tau}(x_0,p_x ,t)\right] \}.
\nonumber\\
\end{eqnarray}
In this expression, $f_P^{\sigma\tau}(x,p_x ,t)$ denotes, in spin-isospin channel
$( \sigma, \tau)$, the average value of reduced Wigner function associated with 
wave functions originating from projectile,
\begin{eqnarray}
\label{eq14}
f_P^{\sigma\tau}(x,p_x ,t) =  \int\!\!\!\int {dydz} \int ds_x \exp \left( - \frac{i}
{\hbar} p_x s_x \right)\nonumber \\
{}\times \sum\limits_{i \in P} \Phi _{i \sigma\tau}^\ast \left(x + \frac{s_x }{2},y,z,t\right)
n_i^{\sigma\tau}\Phi _{i \sigma\tau}\left(x - \frac{s_x }{2},y,z,t\right).\nonumber\\
\end{eqnarray}
The average quantity 
\begin{eqnarray}
\label{eq15}
\bar{f}_P^{\sigma\tau}(x_0,p_x ,t) &=& \frac{f_P^{\sigma\tau}(x_0,p_x,t)}{\Omega(x_0,t)}
\end{eqnarray} 
denotes the reduced Wigner distribution averaged over phase-space 
on the window, i.e., on the plane dividing 
projectile-like and target-like nuclei, where $\Omega(x_0,t)$ is the 
phase-space volume over the window. Quantities $f_T^{\sigma\tau}(x_0,p_x ,t)$ and 
$\bar{f}_T^{\sigma\tau}(x_0,p_x ,t)$ are 
average values of reduced Wigner function associated with wave functions originating 
from target in spin-isospin channel $(\sigma, \tau)$, which are defined
in a similar manner.

\begin{figure}[tbhp]
\begin{center}\leavevmode
\includegraphics[width=\linewidth, clip]{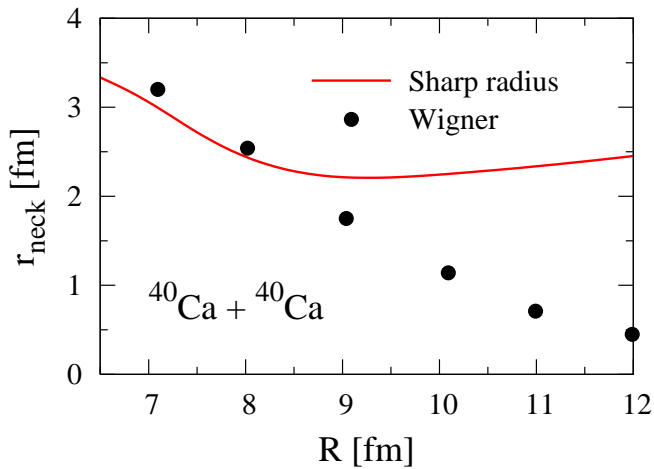}
\caption{
Neck radius determined for the $^{40}$Ca${}+^{40}$Ca reaction at $E_{\rm c.m.}=100$~MeV 
by using Eq.~(\ref{eq17}) (solid line), and by imposing the condition
that the reduced Wigner function $\bar{f}_{P/T}^{\sigma\tau}(x,p_x ,t)$ is close to 1.0 
around the average momentum (filled circles).
}
\label{fig:neckradius}
\end{center}
\end{figure}

We approximate the phase-space volume over the window as
\begin{eqnarray}
\label{eq16}
\Omega(x_0,t)=\pi r_{\rm neck}^{2}(x_0,t)\frac{\pi p_{F}^{2}}{(2\pi\hbar)^2},
\end{eqnarray}
where $p_{F}$ stands for the Fermi momentum.
In this expression $r_{\rm neck}(x_0,t)$ denotes the equivalent sharp radius of the neck, 
which is defined as
\begin{eqnarray}
\label{eq17}
\pi r_{\rm neck}^{2}(x_0,t)=\frac{1}{n_0(x_0,t)} \int dy dz n(x_0,y,z,t),
\end{eqnarray}
where $n(x_0,y,z,t)$ is the local nucleon density while $n_0(x_0,t)$ denotes the density 
at the center of the neck, i.e., 
$n_0(x_0,t) \equiv n(x_0,0,0,t)$. The evolution of $r_{\rm neck}$ deduced from 
Eq.~(\ref{eq17}) is shown by solid lines in Fig.~\ref{fig:neckradius} 
for the $^{40}$Ca${}+^{40}$Ca reaction 
as a function of relative distance. 
While the neck radius has rather reasonable values 
at small $R$, Eq.~(\ref{eq17}) leads to unrealistic large values for well separated nuclei. 
To overcome this difficulty, we use an alternative approach by considering  that 
$\bar{f}_{P/T}^{\sigma\tau}(x_0,p_x ,t)$ should be 
close to 1.0 around the average value of $p_x$. By imposing this condition, 
we directly determine an approximate phase-space volume from Eq.~(\ref{eq15}). 
Then, we deduce $r_{\rm neck}$
at each relative distance $R$ by inverting Eq.~(\ref{eq16}). These are indicated
by filled circles in Fig.~\ref{fig:neckradius}. 
We see that the second prescription not only provides a reasonable 
behavior for $r_{\rm neck}$ at large distances, 
but also matches $r_{\rm neck}$ deduced by using Eq.~(\ref{eq17}) at small distances. 
In the calculations we use the effective neck radius determined by the
second approach. 
\begin{figure}[tbhp]
\begin{center}\leavevmode
\includegraphics[width=\linewidth, clip]{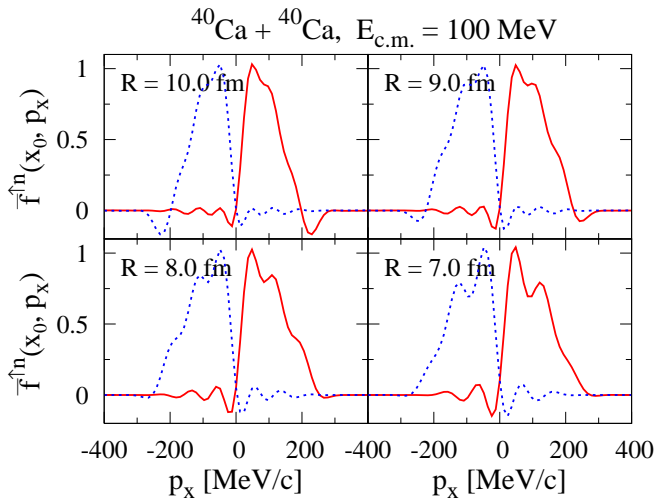}
\caption{
Reduced Wigner function ${\bar f}^{\uparrow n}(x_0,p_x)$
averaged over phase space on the window ($x=x_0$)
for projectile-like (solid line) and target-like (dotted line) nuclei
for the $^{40}$Ca${}+^{40}$Ca reaction at $E_{\rm c.m.}=100$~MeV
at different $R$. The Fermi momentum $p_F$ is taken as 270 MeV/$c$. }
\label{fig:wigner}
\end{center}
\end{figure}
Examples of reduced Wigner function are shown in Fig.~\ref{fig:wigner} for different relative distances. 
Not surprisingly, the reduced Wigner function is sometimes above 1 or below 0. This is indeed expected since
the full quantum Wigner transform is considered without making use of any semiclassical approximation.    

\section{MOMENTUM DIFFUSION COEFFICIENT}

In a recent work~\cite{Washiyama1}, considering simple one-dimensional macroscopic reduction
of TDHF, average transport properties of relative motion in heavy-ion
collisions have been investigated. Temporal evolution of average value of
relative distance $R(t) = \overline {R^\lambda (t)} $ and average value of
relative momentum $P(t) = \overline {P^\lambda (t)} $ are calculated for
average mean-field trajectory determined by the TDHF wave functions.
Relative motion of colliding ions were analyzed in the basis of a simple
classical equation of motion,
\begin{eqnarray}
\label{eq18} 
\frac{d}{dt}P = - \frac{d}{dR}U(R) - \gamma (R)\dot{R}.
\end{eqnarray}
Knowing time evolution of $R(t)$ and $P(t)$, average collective
properties, namely,  nucleus-nucleus potential energy $U(R)$ and  form
factor of one-body friction coefficient $\gamma (R)$ are determined by
inverting Eq.~(\ref{eq18}). In this work, we consider the same geometry of head-on
collision of heavy-ions and extend the macroscopic reduction treatment by considering the
SMF approach. We analyze the relative motion by employing a
Langevin equation. The Langevin equation for the relative motion has the form,
\begin{eqnarray}
\label{eq19} \frac{d}{dt}P^{\lambda} = - \frac{d}{dR^{\lambda}}U(R^{\lambda} ) 
- \gamma (R^{\lambda} )\dot {R}^{\lambda} + \xi_P^{\lambda} (t),
\end{eqnarray}
where $\xi _P^\lambda (t)$ is a Gaussian random force acting on the relative
motion. Ignoring non-Markovian effects, the random force reduces to white
noise specified by a correlation function,
\begin{eqnarray}
\label{eq20}
\overline{\xi _P^\lambda (t)\xi _P^\lambda({t}')} = 2\delta(t-{t}')D_{PP}(R).
\end{eqnarray}
Here $D_{PP}(R)$ denotes the momentum diffusion coefficient, which may
depend on the mean value of the relative distance $R$.

In order to extract momentum diffusion coefficient, we calculate the rate of
change of the relative momentum employing the SMF equations. The
rate of change of relative momentum involves kinetic parts due to nucleon
exchange between projectile and target, and also involves terms arising from
potential energy. In the previous investigation~\cite{Washiyama2}, it is observed that during
evolution from the entrance channel until passing over the Coulomb barrier,
one-body dissipation mechanism is strongly correlated with nucleon exchange
between projectile-like and target-like nuclei. This behavior is similar to
phenomenological nucleon exchange model and the window formula for energy
dissipation~\cite{Randrup2,Feldmeier}. Therefore, in the equation for the 
rate of change of relative momentum, we consider only kinetic terms corresponding to momentum
flow across the window, which can be conveniently expressed in terms of
reduced Wigner distribution as
\begin{eqnarray}
\label{eq21} 
\frac{d}{dt}P^{\lambda} = - \int {\frac{dp_x }{2\pi\hbar }}\frac{p_x^2 }{m}
f^{\lambda} (x,p_x ,t)\vert _{x =x_{0}} +{\rm potential ~ terms}.\nonumber\\
\end{eqnarray}
Small fluctuations of relative momentum are connected to small amplitude
fluctuations in Wigner distribution. Ignoring contribution arising from
potential terms, we have for small fluctuations of relative momentum
\begin{eqnarray}
\label{eq22} 
\frac{d}{dt}\delta P^{\lambda} \approx - \int
{\frac{dp_x }{2\pi \hbar }} \frac{p_x^2 }{m}\delta f^{\lambda}
(x,p_x ,t)\vert _{x =x_{0}} = \xi _P^{\lambda} (t).
\end{eqnarray}
The right hand side in this expression acts as a random force for generating
fluctuations in the relative momentum. Since $\delta f^\lambda (x,p_x ,t)$
is a Gaussian random quantity, the random force $\xi _P (t)$ is also
Gaussian random, which is specified by a correlation function,
\begin{eqnarray}
\label{eq23} \overline {\xi _P^\lambda (t)\xi _P^\lambda ({t}')}& =&
\int\!\!\!\int \frac{dp_x }{2\pi \hbar}\frac{d{p}^{\prime}_x}{2\pi\hbar}
\frac{p_x^2}{m}\frac{{p'}_x^2}{m}\nonumber\\
&&{}\times\overline{\delta f^{\lambda}(x,p_x,t)
\delta f^{\lambda}(x,{p}^{\prime}_x ,{t}')}\vert _{x = x_{0}}.
\end{eqnarray}
Using the expression for the correlation function of the reduced Wigner
distribution in Eq.~(\ref{eq12}), according to Eq.~(\ref{eq20}), 
the momentum diffusion coefficient is given by
\begin{eqnarray}
\label{eq24}
D_{PP}(t) =\int \frac{dp_x }{2\pi \hbar} \frac{\vert p_x \vert}{m}
\frac{p_x^2}{2}\Lambda^{+}(x_{0},p_x ,t).
\end{eqnarray}
From the SMF approach, we cannot directly derive an expression for the
friction coefficient $\gamma (R)$. The reason is that we cannot associate
the net momentum flow across the window, which is given by the first term on
the right side of Eq.~(\ref{eq21}), with dissipative force acting on the relative
motion. However, from the expression~(\ref{eq24}) for diffusion coefficient and from
the random walk mechanism of nucleon exchange~\cite{Randrup2,Feldmeier}, 
we can infer an expression for the friction coefficient. In the expression for diffusion
coefficient, first and second terms correspond to nucleon flux from
projectile to target and from target to projectile, respectively. Each
nucleon transfer changes the relative momentum by an amount $p_x $ and
increases the dispersion of the relative momentum by an amount $p_x^2 $.
Nucleon transfer in both direction increases dispersion of relative
momentum, therefore diffusion coefficient is determined by total nucleon
flux, i.e., sum of flux from projectile to target and from target to
projectile. On the other hand, dissipation is determined by the net momentum
flow through the window. Hence, the resultant dissipative force can be expressed as
\begin{eqnarray}
\label{eq25}
F(t) = \int{\frac{dp_x }{2\pi \hbar }} \frac{\vert p_x \vert}{m} p_x 
\Lambda^{-}(x_{0},p_x ,t).
\end{eqnarray}
Then, it is possible to deduce from TDHF simulations the momentum diffusion
coefficient $D_{PP}(t)=D_{PP}(R)$ and the friction force $F(t)$ as a function
of relative distance. We note that these transport coefficients correspond to 
the phenomenological window formula arising from 
the nucleon exchange mechanism~\cite{Feldmeier}, and they are determined
in terms of the average evolution specified by the TDHF.

\begin{figure}[tbhp]
\begin{center}\leavevmode
\includegraphics[width=\linewidth, clip]{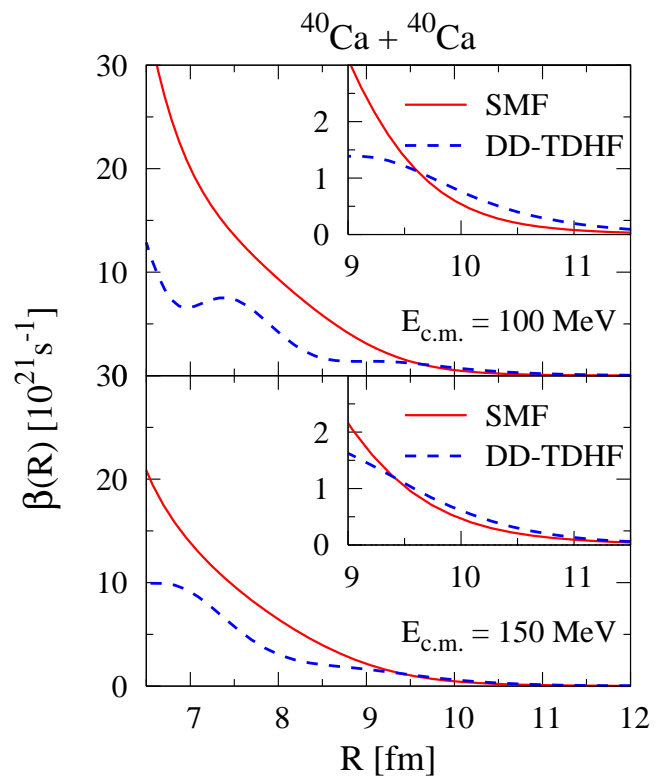}
\caption{
Reduced friction coefficient $\beta(R) =\gamma(R)/\mu$ as a function of $R$
for the $^{40}$Ca${}+^{40}$Ca reaction at $E_{\rm c.m.}=100$~MeV (upper panel)
and at $E_{\rm c.m.}=150$~MeV (lower panel).
For each energy, a zoom on the Coulomb barrier region is also shown in the insert.
}
\label{fig:friction}
\end{center}
\end{figure}

Rather than calculating the dissipative force, it is more instructive to calculate 
the friction coefficient $\gamma(R)$. For this purpose, we assume that dissipative 
force is proportional to relative velocity, i.e., $F(t) = -\gamma (R)\dot {R}$, and consider 
the reduced friction coefficient $\beta(R) =\gamma(R)/\mu(R)$, where $\mu(R)$ denotes 
inertia associated with relative motion. Solid lines in Fig.~\ref{fig:friction} 
show the reduced friction coefficient as a function of $R$ for head-on collision of 
$^{40}$Ca${}+^{40}$Ca at two different center-of-mass energies. For each energy, enlarged
plot around the Coulomb barrier region is shown in the insert. In a recent work~\cite{Washiyama2}, 
we extracted the reduced friction coefficient associated with relative motion
employing a different reduction procedure, so-called Dissipative-Dynamics TDHF (DD-TDHF),
which, in principle, incorporates dissipation due to both window and wall mechanisms. 
Dashed lines in Fig.~\ref{fig:friction} show the results of this reduction procedure.
Good agreement is found between two different calculations above and close to the Coulomb 
barrier ($\sim 9.8$~fm).  Below the Coulomb barrier, the DD-TDHF method is not reliable.
However, the method based on the SMF provides a proper description of the one-body friction 
coefficient due to nucleon exchange mechanism for a wide range of relative distance. 

\begin{figure}[tbhp]
\begin{center}\leavevmode
\includegraphics[width=\linewidth, clip]{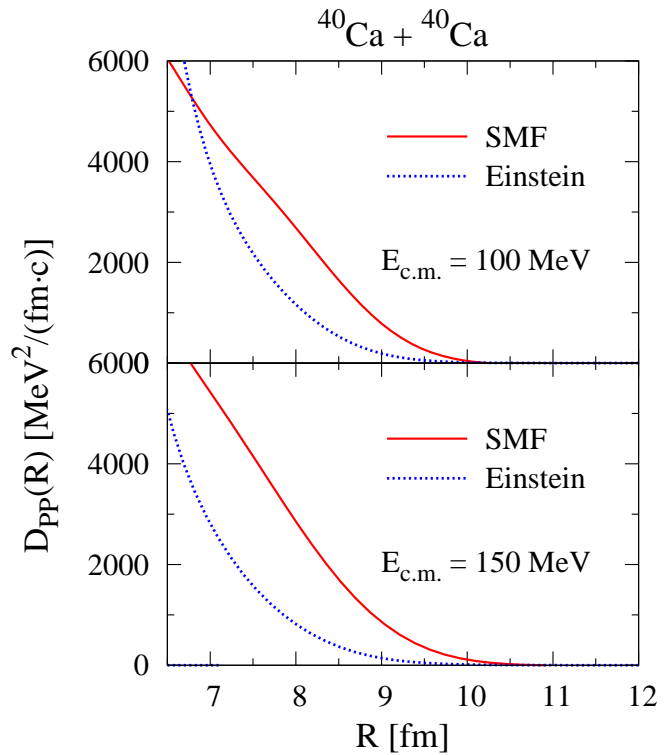}
\caption{
Diffusion coefficient $D_{PP}$ obtained by SMF (solid lines)
and by the Einstein relation $D_{PP}^{\rm eq}=\gamma(R) T(t)$ (dashed lines)
as a function of $R$ for the $^{40}$Ca${}+^{40}$Ca reaction at $E_{\rm c.m.}=100$~MeV 
(upper panel) and at $E_{\rm c.m.}=150$~MeV (lower panel).}
\label{fig:diffusion}
\end{center}
\end{figure}

Solid lines in Fig.~\ref{fig:diffusion} show the momentum diffusion coefficient $D_{PP}$, 
Eq.~(\ref{eq24}), as a function of $R$ for head-on collision of $^{40}$Ca${}+^{40}$Ca at 
two different center-of-mass energies. 
Similarly to the reduced friction coefficient, 
magnitude of the momentum diffusion coefficient increases for decreasing relative distance. 
The increase of magnitude of transport coefficients, i.e., 
friction and diffusion coefficients, for decreasing $R$ is essentially due to 
larger window area and larger number of nucleon exchange between projectile-like and 
target-like nuclei. It is important to realize that, even though 
the ordinary TDHF does not contain information 
about density fluctuations, we can employ the average information provided by the TDHF to 
calculate diffusion coefficients associated with macroscopic variables. In practical applications,
the momentum diffusion coefficient is usually taken as the thermal equilibrium value determined
by the Einstein relation in terms of friction coefficient and effective temperature as
\begin{eqnarray}
\label{eq26}
D^{\rm eq}_{PP}(R)=\gamma(R)T(t)
\end{eqnarray} 
In this expression, $T(t)$ denotes the effective temperature assuming local equilibrium.
It can be determined in terms of
excitation energy denoted by $E^*$ by the relation $T(t)=\sqrt{E^{*}(t)/a}$, where $a$ denotes 
level density parameter, taken here as $a=A/12$. We can estimate the excitation energy
in terms of dissipated energy according to
\begin{eqnarray}
\label{eq27}
E^{*}(t)=\int_{0}^{t}dt^{\prime}\gamma [R(t^{\prime})][\dot{R}(t^{\prime})]^{2}
\end{eqnarray} 
Dashed lines in Fig.~\ref{fig:diffusion} show the diffusion coefficient 
$D^{\rm eq}_{PP}(R)$ determined according to the Einstein relation. As seen from
the figure, the Einstein relation severely underestimates magnitude of
dynamical diffusion coefficient. The fact that the Einstein relation severely 
underestimates the dynamical diffusion coefficient associated with the relative 
motion was already realized in the phenomenological description of nucleon 
exchange model in Ref.~\cite{Feldmeier}. 

\section{CONCLUSIONS}

Recently proposed stochastic mean-field
theory incorporates both one-body dissipation and fluctuation
mechanisms in a manner consistent with quantal
fluctuation-dissipation theorem of non-equilibrium statistical
mechanics~\cite{Ayik2}. This was illustrated for slow collective motion by projecting
equation of motion of the SMF onto a collective space in adiabatic limit.
The projection gives rise to a generalized Langevin equation for collective variables, in
which mean-field dissipation and fluctuation mechanisms are connected through 
the quantal fluctuation-dissipation relation. Therefore, this approach provides a powerful tool for
microscopic description of low energy nuclear processes in which
two-body dissipation and fluctuation mechanisms do not play
important role. The low energy processes include induced fission,
heavy-ion fusion near barrier energies, and spinodal decomposition 
during the expansion phase of hot piece of nuclear matter produced 
in heavy-ion collisions~\cite{Ayik3,Maria}. 

In this work, we carry out a different projection of the SMF approach on the relative motion
in fusion reaction by following the DD-TDHF method introduced in~\cite{Washiyama1}
and deduce one-body friction and one-body diffusion coefficients associated with 
relative motion.  It is remarkable that expressions of transport coefficients
for the relative motion (as well as transport coefficients for other macroscopic 
variables which are not mentioned in this work) have the same form as given 
by the phenomenological nucleon exchange model~\cite{Randrup2,Feldmeier}. 
The phenomenological nucleon exchange model involves an important assumption, namely, when
a nucleon passes through the window, it instantaneously equilibrates with the new environment
on the other side of the window. On the other hand, transport coefficients deduced 
from the SMF approach
do not involves this assumption, and also they are not restricted 
by adiabatic or diabatic approximation. Therefore, these transport coefficients provide 
a microscopic basis for determining magnitude of the actual one-body dissipation and the corresponding 
mean-field fluctuation mechanism.  We also stress the fact that, assuming amplitude of 
density fluctuations are small, 
transport coefficients are calculated in terms of the average evolution determined by TDHF simulations 
as a function of relative distance. In the continuation of the investigations, 
we plan to generalize the projection
procedure of the SMF approach for off-central collisions and 
also deduce transport coefficients for nucleon diffusion in grazing heavy-ion collisions.

\begin{acknowledgments}
We thank P. Bonche for providing the 3D-TDHF code.
S.A. gratefully acknowledges the CNRS for financial support and
GANIL for warm hospitality extended to him during his visit. 
K.W. gratefully acknowledges the French Embassy 
in Japan for financial supports.
This work is supported in part by the US DOE Grant No. DE-FG05-89ER40530.
\end{acknowledgments}


\appendix*
\section{CORRELATION FUNCTION OF WIGNER DISTRIBUTION}
Small amplitude fluctuations of Wigner distribution can be expressed as
\begin{eqnarray}
\label{eqA.1}
&&\delta f^\lambda (\boldsymbol {r},\boldsymbol {p},t) = \int d^3s\exp \left( - \frac{i}{\hbar} 
\boldsymbol {p} \cdot \boldsymbol {s}\right)\nonumber\\
&&{}\times\sum\limits_{ij \sigma\tau} 
\Phi_{j \sigma\tau}^\ast \left(\boldsymbol {r} +\frac{\boldsymbol {s}}{2} ,t\right)
\delta \rho_{ij}^\lambda(\sigma\tau)
\Phi_{i \sigma\tau}\left(\boldsymbol {r} -\frac{\boldsymbol {s}}{2},t\right),\nonumber\\
\end{eqnarray}
where the single-particle wave functions are complete set of solutions of
the ordinary TDHF. The initial values of stochastic expansion
coefficients $\delta \rho _{ij}^\lambda(\sigma\tau) $ are Gaussian random numbers as specified by
Eq.~(\ref{eq4}). In principle, these coefficients evolve in time according to
time-dependent RPA equations. Here, we ignore this evolution and take them
as Gaussian random numbers as specified by the initial conditions.
Fluctuating part of the density matrix can be separated into four
groups, $\delta \rho _{PP}^\lambda $,
$\delta \rho _{TT}^\lambda $, $\delta \rho _{PT}^\lambda $ and 
$\delta \rho_{TP}^\lambda $, which are associated with wave functions
originating from projectile and target nuclei and the mixed terms. As a
result, small amplitude fluctuations of the Wigner distribution separate
into four parts, $\delta f_{PP}^\lambda (\boldsymbol {r},\boldsymbol {p},t)$, $\delta
f_{TT}^\lambda (\boldsymbol {r},\boldsymbol {p},t)$, $\delta f_{PT}^\lambda (\boldsymbol {r},\boldsymbol{p},t)$ 
and $\delta f_{TP}^\lambda (\boldsymbol {r},\boldsymbol {p},t)$. We calculate the
equal time correlation function of the Wigner distribution in semi-classical
approximation. First, we consider the correlation function associated with 
wave functions originating from projectile. Using the expression~(\ref{eq4}) 
for the variance of the matrix elements, we deduce
\begin{widetext}\begin{eqnarray}
\label{eqA.2}
&&\overline {\delta f_{PP}^\lambda (\boldsymbol {r},\boldsymbol {p},t)
\delta f_{PP}^\lambda({\boldsymbol {r}}^{\prime},{\boldsymbol {p}}^{\prime},t)}
= \int\!\!\!\int {d^3sd^3{s}'\exp \left( -\frac{i}{\hbar }\boldsymbol {p} \cdot \boldsymbol {s} \right)} 
\exp \left( - \frac{i}{\hbar }{\boldsymbol{p}}^{\prime} \cdot {\boldsymbol {s}}^{\prime} \right)   \nonumber\\
&&{}\times\sum\limits_{ij \sigma\tau\in P} {\Phi _{j \sigma\tau}^\ast 
\left(\boldsymbol {r} + \frac{\boldsymbol {s}}{2},t \right) }
\Phi_{i \sigma\tau} \left( \boldsymbol {r} - \frac{\boldsymbol {s}}{2},t \right)
\Phi _{i \sigma\tau}^\ast \left( {\boldsymbol {r}}^{\prime} + 
\frac{{\boldsymbol{s}}^{\prime}}{2},t \right) 
\Phi _{j \sigma\tau} \left( {\boldsymbol {r}}^{\prime} - 
\frac{{\boldsymbol {s}}^{\prime}}{2},t\right)n_j^{\sigma\tau}(1 - n_i^{\sigma\tau}).
\end{eqnarray}
In the term that is proportional to $n_j^{\sigma\tau} $, we use the closure relations to find,
\begin{eqnarray}
\label{eqA.3}
\sum\limits_{i \in P} {\Phi _{i \sigma\tau}^\ast \left({\boldsymbol {r}}^{\prime} + 
\frac{{\boldsymbol {s}}^{\prime}}{2},t \right) } \Phi _{i \sigma\tau} \left(
\boldsymbol {r} - \frac{\boldsymbol {s}}{2},t \right) = 
\delta \left( {\boldsymbol {r}}^{\prime} - \boldsymbol{r} + \frac{{\boldsymbol {s}}^{\prime} + \boldsymbol {s}}{2} \right).
\end{eqnarray}
In this expression, summation runs over a complete set of single-particle
states, i.e., occupied and unoccupied states originating from the projectile. The
closure relation satisfied by the complete set of states at the initial
state will remain valid at later times. In the second step, we introduce the
Wigner distribution,
\begin{eqnarray}
\label{eqA.4}
\sum\limits_{j \in P} \Phi _{j \sigma\tau}^\ast \left(\boldsymbol {r} + \frac{\boldsymbol {s}}{2} ,t\right)
n_j^{\sigma\tau}
\Phi _{j \sigma\tau} \left({\boldsymbol {r}}^{\prime} - \frac{{\boldsymbol {s}}^{\prime}}{2},t\right)
=\int \frac{d^3Q}{(2\pi \hbar )^3}\exp \left[\frac{i}{\hbar }\left(\boldsymbol {r} - {\boldsymbol{r}}^{\prime} 
+ \frac{\boldsymbol {s} + {\boldsymbol {s}}^{\prime}}{2} \right) \cdot \boldsymbol {Q}\right]
 f_P^{\sigma\tau} \left(\frac{\boldsymbol{r} + {\boldsymbol {r}}^{\prime}}{2} + 
\frac{\boldsymbol {s} - {\boldsymbol {s}}^{\prime}}{4},\boldsymbol {Q},t\right), \nonumber\\
\end{eqnarray}
where $f_P^{\sigma\tau} (\boldsymbol {r},\boldsymbol {p},t)$ denotes the ensemble averaged Wigner
distribution associated with wave functions originating from projectile 
in spin-isospin channel $(\sigma,\tau)$.
After making transformations, $\boldsymbol {s} = {}+ \boldsymbol {\eta }+\boldsymbol {Y}/2$ 
and ${\boldsymbol {s}}^{\prime} ={} - \boldsymbol {\eta } + \boldsymbol {Y} / 2$, 
the term that is proportional to occupation factor $n_j^{\sigma\tau}$ 
in the right-hand side of Eq.~(\ref{eqA.2}) becomes
\begin{eqnarray}
\label{eqA.5}
&&\sum_{\sigma\tau}\int\!\!\!\int\!\!\!\int d^3Yd^3\eta \exp \left[ - \frac{i}{\hbar}\left(\boldsymbol {Y} 
\cdot \frac{\boldsymbol {p} + {\boldsymbol {p}}^{\prime}}{2} \right)\right]
\exp \left[ - \frac{i}{\hbar}\boldsymbol {\eta } \cdot (\boldsymbol {p} - {\boldsymbol {p}}^{\prime})\right]
\nonumber\\&&{}\times 
\frac{d^3Q}{(2\pi \hbar )^3}\exp \left[\frac{i}{\hbar }\boldsymbol {Y} \cdot \boldsymbol{Q}\right]  
f_P^{\sigma\tau} \left(\boldsymbol {r} - \frac{\boldsymbol {Y}}{4} + \frac{\boldsymbol {\eta }}{2},\boldsymbol{Q},t\right)
\delta \left({\boldsymbol {r}}^{\prime}-\boldsymbol{r} + \frac{\boldsymbol {Y}}{2}\right).
\end{eqnarray}
Assuming that the Wigner distribution is a smooth function of $\boldsymbol {r}$,
$f_P^{\sigma\tau}(\boldsymbol {r} - \frac{\boldsymbol {Y}}{4} + \frac{\boldsymbol {\eta }}{2},\boldsymbol {Q},t)
\approx f_P^{\sigma\tau}(\boldsymbol {r},\boldsymbol {Q},t)$ and $\delta ({\boldsymbol {r}}^{\prime} -\boldsymbol{r} +
\frac{\boldsymbol {Y}}{2}) \approx \delta (\boldsymbol {r} - {\boldsymbol {r}}^{\prime})$, we can carry
out the integrations over $\boldsymbol {\eta }$ and $\boldsymbol {Y}$ to obtain 
$(2\pi\hbar )^3\delta (\boldsymbol {p} - {\boldsymbol {p}}')$ and $(2\pi\hbar)^3\delta(\boldsymbol {p} - \boldsymbol{Q})$, 
respectively. As a result, the term (\ref{eqA.5}) becomes 
\begin{eqnarray}
\label{eqA.6}
{\rm (A.5)} = (2\pi \hbar )^3\delta (\boldsymbol {p} -\boldsymbol{p}^{\prime})
\delta (\boldsymbol {r} - {\boldsymbol {r}}^{\prime})\sum\limits_{\sigma\tau}
f_P^{\sigma\tau} (\boldsymbol {r},\boldsymbol {p},t).
\end{eqnarray}
%
%
For the term proportional to $n_i^{\sigma\tau}n_j^{\sigma\tau}$ in Eq.~(\ref{eqA.2}),
again we introduce the Wigner distribution for the factor involving the index $j$,
\begin{eqnarray}
\label{eqA.7}
\sum\limits_{j \in P} \Phi _{j\sigma\tau}^\ast \left(\boldsymbol {r} + \frac{\boldsymbol{s}}{2}\right) 
n_j^{\sigma\tau} \Phi_{j\sigma\tau} \left( {\boldsymbol {r}}^{\prime} - \frac{ {\boldsymbol {s}}^{\prime} }{2}\right)
=\int \frac{d^3Q_1 }{(2\pi \hbar )^3}\exp \left[\frac{i}{\hbar }\left(\boldsymbol {r} -
{\boldsymbol {r}}^{\prime} + \frac{\boldsymbol {s} + \boldsymbol {s}^{\prime}}{2}\right) 
\cdot \boldsymbol {Q}_1 \right]
f_P^{\sigma\tau}\left(\frac{\boldsymbol {r} + {\boldsymbol {r}}^{\prime}}{2} 
+ \frac{\boldsymbol {s} - {\boldsymbol {s}}^{\prime}}{4},\boldsymbol{Q}_1 ,t\right),\nonumber\\
\end{eqnarray}
and for the one involving the index $i$,
\begin{eqnarray}
\label{eqA.8}
\sum\limits_{i \in P} \Phi_{i \sigma\tau}^\ast 
\left({\boldsymbol {r}}^{\prime} + \frac{{\boldsymbol{s}}^{\prime}}{2}\right) 
n_i^{\sigma\tau} 
\Phi_{i \sigma\tau}\left(\boldsymbol {r} - \frac{\boldsymbol {s}}{2}\right)
=\int \frac{d^3Q_2 }{(2\pi \hbar )^3}
\exp \left[ \frac{i}{\hbar }
\left({\boldsymbol {r}}^{\prime} -\boldsymbol {r} + \frac{\boldsymbol {s} + {\boldsymbol {s}}^{\prime}}{2} \right) 
\cdot \boldsymbol {Q}_2 \right]
f_P^{\sigma\tau}\left(\frac{\boldsymbol {r} + {\boldsymbol {r}}^{\prime}}{2} + \frac{{\boldsymbol {s}}^{\prime} - \boldsymbol {s}}{4},\boldsymbol{Q}_2 ,t\right).\nonumber\\
\end{eqnarray}
Making the same transformations, $\boldsymbol {s} = + \boldsymbol {\eta } + \boldsymbol {Y} /2$ 
and ${\boldsymbol {s}}^{\prime} = - \boldsymbol {\eta } + \boldsymbol {Y} / 2$, 
the term that is proportional to $n_i^{\sigma\tau}n_j^{\sigma\tau}$ 
in the right-hand side of Eq.~(\ref{eqA.2}) becomes
\begin{eqnarray}
\label{eqA.9}
&&\sum_{\sigma\tau} \int\!\!\!\int\!\!\! \int\!\!\!\int d^3Yd^3\eta 
\exp\left[-\frac{i}{\hbar}\left(\boldsymbol {Y}\cdot \frac{\boldsymbol{p} + {\boldsymbol{p}}^{\prime}}{2} \right)\right]
\exp\left[-\frac{i}{\hbar}\boldsymbol {\eta } \cdot (\boldsymbol {p} - {\boldsymbol {p}}^{\prime})\right]
\frac{d^3Q_1 }{(2\pi \hbar )^3} \frac{d^3Q_2}{(2\pi \hbar )^3}
\exp \left[\frac{i}{\hbar }
\left(\boldsymbol {r} - {\boldsymbol {r}}^{\prime} + \frac{\boldsymbol{Y}}{2}\right) \cdot \boldsymbol {Q}_1 \right] 
\nonumber \\&&{}\times 
\exp \left[\frac{i}{\hbar }
\left({\boldsymbol {r}}^{\prime} - \boldsymbol {r} + \frac{\boldsymbol {Y}}{2}\right) \cdot \boldsymbol {Q}_2 \right]
f_P^{\sigma\tau} 
\left(\frac{\boldsymbol {r} + {\boldsymbol {r}}^{\prime}}{2} + \frac{\boldsymbol {\eta }}{2},\boldsymbol {Q}_1,t\right)
f_P^{\sigma\tau } 
\left(\frac{\boldsymbol {r} + {\boldsymbol {r}}^{\prime}}{2} - \frac{\boldsymbol {\eta }}{2},\boldsymbol {Q}_2,t\right).
\end{eqnarray}
We introduce another change of variables $\boldsymbol {Q}_1 = \boldsymbol {Q} + \boldsymbol {q} /2$ 
and $\boldsymbol {Q}_2 = \boldsymbol {Q} - \boldsymbol {q} / 2$, and again assume that the Wigner
distribution has a smooth function of $\boldsymbol {r}$ and ignore $\boldsymbol {\eta }$
dependence. Then, integrations over $\boldsymbol {\eta }$, $\boldsymbol {q}$ and $\boldsymbol
{Y}$ give $(2\pi \hbar )^3\delta (\boldsymbol {p} - {\boldsymbol {p}}^{\prime})$, 
$(2\pi \hbar )^3\delta (\boldsymbol {r}- {\boldsymbol {r}}^{\prime})$ 
with $\boldsymbol {Q}_1 \approx \boldsymbol {Q}_2 = \boldsymbol {Q}$
and $(2\pi \hbar )^3\delta (\boldsymbol {p}- {\boldsymbol {Q}})$,
respectively. As a result, the term~(\ref{eqA.9}) becomes
\begin{eqnarray}
\label{eqA.10}
{\rm (A.9)}=(2\pi \hbar )^3\delta (\boldsymbol {p} - {\boldsymbol {p}}^{\prime})
\delta (\boldsymbol {r}-{\boldsymbol {r}}^{\prime})
 \sum\limits_{\sigma\tau}f_P^{\sigma\tau}(\boldsymbol {r},\boldsymbol {p},t)
 f_P^{\sigma\tau} (\boldsymbol {r},\boldsymbol {p},t).
\end{eqnarray}
Combining together, equal time correlation function (\ref{eqA.2}) of the Wigner
distribution associated with wave functions originating from projectile
becomes,
\begin{eqnarray}
\label{eqA.11}
\overline{\delta f_{PP}^\lambda (\boldsymbol {r},\boldsymbol {p},t)\delta f_{PP}^\lambda
({\boldsymbol {r}}^{\prime},{\boldsymbol {p}}^{\prime},t)}
=(2\pi \hbar )^3\delta (\boldsymbol {p} - {\boldsymbol{p}}^{\prime})
\delta (\boldsymbol {r} - {\boldsymbol {r}}^{\prime})
\sum\limits_{\sigma\tau}f_P^{\sigma\tau} (\boldsymbol {r},\boldsymbol {p},t)
[1 - f_P^{\sigma\tau} (\boldsymbol{r},\boldsymbol {p},t)].
\end{eqnarray}
In a similar manner, we can calculate the correlation function of the Wigner
distribution associated with wave functions originating from target and from
mixed configuration,
\begin{eqnarray}
\label{eqA.12}
\overline{\delta f_{TT}^\lambda (\boldsymbol {r},\boldsymbol {p},t)\delta f_{TT}^\lambda
({\boldsymbol {r}}^{\prime},{\boldsymbol {p}}^{\prime},t)}
=(2\pi \hbar )^3\delta (\boldsymbol {p} - {\boldsymbol{p}}^{\prime})
\delta (\boldsymbol {r} - {\boldsymbol {r}}^{\prime})
 \sum\limits_{\sigma\tau}f_T^{\sigma\tau} (\boldsymbol {r},\boldsymbol {p},t)
\left[1 - f_T^{\sigma\tau} (\boldsymbol{r},\boldsymbol {p},t)\right]
\end{eqnarray}
and
\begin{eqnarray}
\label{eqA.13}
\overline {\delta f_{PT}^\lambda (\boldsymbol {r},\boldsymbol {p},t)\delta f_{PT}^\lambda
({\boldsymbol {r}}^{\prime},{\boldsymbol {p}}^{\prime},t)} 
=(2\pi \hbar )^3\delta (\boldsymbol {p} - {\boldsymbol{p}}^{\prime})
\delta (\boldsymbol {r} - {\boldsymbol {r}}^{\prime})\Lambda^{+}(\boldsymbol {r},\boldsymbol {p},t),
\end{eqnarray}
where
\begin{eqnarray}
\label{eqA.14}
\Lambda^{+}(\boldsymbol {r},\boldsymbol {p},t)=\sum\limits_{\sigma\tau} 
\left\{f_P^{\sigma\tau}(\boldsymbol {r},\boldsymbol {p},t)
\left[1 - f_T^{\sigma\tau}(\boldsymbol {r},\boldsymbol {p},t)\right] 
+ f_T^{\sigma\tau}(\boldsymbol {r},\boldsymbol {p},t)\left[1 - f_P^{\sigma\tau}(\boldsymbol {r},\boldsymbol {p},t)\right] \right\}.
\end{eqnarray}
Total correlation function of the Wigner distribution is the sum of (\ref{eqA.11}),
(\ref{eqA.12}) and (\ref{eqA.13}). In the mean-field description, the sub-spaces of wave
functions originating from projectile and target nuclei behave like pure
states. Therefore, contributions of correlations coming from direct terms involving
$f_P^{\sigma\tau}(\boldsymbol {r},\boldsymbol {p},t)\left[1 - f_P^{\sigma\tau}(\boldsymbol {r},\boldsymbol {p},t)\right]$ and 
$f_T^{\sigma\tau}(\boldsymbol{r},\boldsymbol {p},t)\left[1 - f_T^{\sigma\tau}(\boldsymbol {r},\boldsymbol {p},t)\right]$ are 
expected to be small. Hence, we can approximately express the total correlation function of Wigner
distribution as,
\begin{eqnarray}
\label{eqA.15}
\overline {\delta f^\lambda (\boldsymbol {r},\boldsymbol {p},t)\delta f^\lambda 
({\boldsymbol{r}}^{\prime},{\boldsymbol {p}}^{\prime},t)}
\approx(2\pi \hbar )^3\delta (\boldsymbol {p} - {\boldsymbol{p}}^{\prime})
\delta (\boldsymbol {r} - {\boldsymbol {r}}^{\prime})\Lambda^{+}(\boldsymbol {r},\boldsymbol {p},t). 
\end{eqnarray}

We also want to calculate different time correlation function of the
Wigner distribution. Assuming that the correlation function has short correlation
time, i.e., much shorter than mean-free path, different time correlation
function can be deduced by observing that in short time intervals of order
of correlation time $\vert t - {t}'\vert \le \tau _{corr} $, Wigner
distribution may be approximated as a free propagation, $\delta f(\boldsymbol{r},\boldsymbol {p},t + \tau ) 
\approx \delta f(\boldsymbol {r} - \tau \boldsymbol {p} / m,\boldsymbol{p},t)$. 
As a result, different time correlation function can be expressed as,
\begin{eqnarray}
\label{eqA.16}
\overline {\delta f^\lambda (\boldsymbol {r},\boldsymbol {p},t)\delta f^\lambda ({\boldsymbol{r}}^{\prime},
{\boldsymbol {p}}^{\prime},{t}')}
= (2\pi \hbar )^3\delta (\boldsymbol {p} - {\boldsymbol{p}}^{\prime})
\delta \left[\boldsymbol {r} - {\boldsymbol {r}}^{\prime} - (t - {t}')\boldsymbol {p}/m\right]
\Lambda^{+}(\boldsymbol {r},\boldsymbol {p},t).
\end{eqnarray}
In order to deduce the correlation function on the window, $x = {x}'= x_0$, we notice that
\begin{eqnarray}
\label{eqA.17}
\delta \left[\boldsymbol {r} - {\boldsymbol {r}}^{\prime} - (t - {t}')\boldsymbol{p}/m\right]
 \to\frac{m}{\vert
p_x \vert }\delta (t - {t}')\delta (y - {y}')\delta (z - {z}').
\end{eqnarray}
In determining transport coefficients, we need to carry out integration over window variables, 
$y,z,p_y ,p_z $, of product of Wigner distributions. Since construction of three-dimensional 
Wigner functions in terms of TDHF wave functions requires a large numerical effort, 
we introduce the following approximation for the phase-space integration over the window, 
\begin{eqnarray}
\label{eqA.18}
\int\!\!\!\int {dydz\frac{dp_y dp_z }{(2\pi \hbar )^2}}
f_P^{\sigma\tau}(\boldsymbol {r},\boldsymbol{p},t)f_T^{\sigma\tau}(\boldsymbol {r},\boldsymbol {p},t) 
\approx \frac{1}{\Omega(x,t)} f_P^{\sigma\tau}(x,p_x ,t)f_T^{\sigma\tau}(x,p_x ,t).
\end{eqnarray}
\end{widetext}
Here $\Omega(x,t)$ denotes the phase-space volume on the window. As a result, the
correlation function on the window can be expressed in terms of the reduced Wigner distributions 
along $x$-axis given by Eq.~(\ref{eq12}).

\end{document}